\documentclass{emulateapj}
\usepackage[usenames]{color}
\usepackage{amsmath}

\usepackage{rotating}

\shorttitle{Magnetic field in high-mass IRDCs}
\shortauthors{Pillai et al.}

\begin{document}

\title{Magnetic fields in High-Mass Infrared Dark Clouds}

\author{T.\ Pillai \altaffilmark{1,2}, J.\ Kauffmann
  \altaffilmark{1,2}, J.C.\ Tan \altaffilmark{3}, P.F.\ Goldsmith
  \altaffilmark{4}, S.J.\ Carey \altaffilmark{5}, K.M.\ Menten \altaffilmark{2}}
\email{tpillai.astro@gmail.com}

\altaffiltext{1}{California Institute of Technology, Cahill Center for
  Astronomy and Astrophysics, Pasadena, CA 91125, USA}
\altaffiltext{2}{Max Planck Institut f\"ur Radioastronomie}
\altaffiltext{3}{University of Florida}
\altaffiltext{4}{Jet Propulsion Laboratory, California Institute of Technology}
\altaffiltext{5}{IPAC, California Institute of Technology}

%\slugcomment{Copyright 2012 California Institute of Technology. All
%  rights reserved. Government sponsorship acknowledged.}

\begin{abstract}
High-mass Stars are cosmic engines known to dominate the energetics
in the Milky Way and other galaxies. However, their formation is still
not well understood. Massive, cold, dense clouds, often appearing as
Infrared Dark Clouds (IRDCs), are the nurseries of massive stars. No
measurements of magnetic fields in IRDCs in a state prior to the onset
of high-mass star formation (HMSF) have previously been available,
and prevailing HMSF theories do not consider strong magnetic
fields. Here, we report observations of magnetic fields in two of the
most massive IRDCs in the Milky Way. We show that IRDCs G11.11$-$0.12
and G0.253$+$0.016 are strongly
magnetized and that the strong magnetic field is as important as
turbulence and gravity for HMSF. The main dense filament in G11.11$-$0.12 is perpendicular to the
magnetic field, while the lower density filament merging onto the main
filament is parallel to the magnetic
field. The implied magnetic field is strong enough to
suppress fragmentation sufficiently to allow HMSF. Other mechanisms
reducing fragmentation, such as the entrapment of heating from young
stars via high mass surface densities, are not required to facilitate
HMSF.
\end{abstract}


\maketitle

\section{Introduction}
\label{sec:introduction}
High-mass (O and B type, $M>8$\, $\hbox{M}_\odot$) stars live wild and die
young. In spite of their short life time, they dominate the energetics
over a wide range of scales from the interstellar medium of the Milky
Way to that of high-redshift galaxies.  While these stars
play a crucial role in the cosmos, there are several aspects of their
formation that are still not well understood \citep{tan2014}.

The birth sites of massive stars, cold dense cores, can often appear
as IRDCs, silhouetted against the diffuse MIR emission of the Galactic
plane \citep{perault1996:iso,carey1998:irdc}. The evolution of these natal clouds is
expected to be governed by some combination of gravity, turbulence and
magnetic fields, but the relative importance of these effects has been
difficult to determine. While the first two factors can be explored
relatively easily, the signatures of magnetic fields are much harder
to detect.  No estimate of magnetic fields strengths in IRDCs in a state
prior to the onset of HMSF have previously
been available \citep{tan2014}.  

The IRDCs G11.11$-$0.12  and G0.253$+$0.016 are among the first discovered and
darkest shadows in the galactic plane \citep{carey1998:irdc}. G11.11$-$0.12 is at a distance of
3.6\,kpc, has a length $\sim{}30$\,pc, a mass of $10^5\,M_{\odot}$, and
is known to host a single site with a high-mass
protostar \citep{pillai2006a:g11,henning2010:g11,kainulainen2013}. G0.253$+$0.016 with 9\,pc length and
$10^5\,M_{\odot}$ lies in the Galactic Center region at $\approx{}8.4$\,kpc
distance; it is one of the most massive and dense clouds in the Galaxy
but does not yet host high-mass stars \citep{lis_menten1998:irdc,longmore2012:m0.25,kauffmann2013:gc}.
Largely unperturbed by active high-mass star formation, these two clouds thus
provide ideal sites for studying the magnetic field at the very
beginning of high-mass star formation.

Polarized thermal dust emission can trace the magnetic field in
molecular clouds.  Dust grains in molecular clouds become aligned with their major axes preferentially oriented perpendicular to the magnetic field most likely through radiative torques \citep{lazarian2007}. This mechanism of grain alignment requires asymmetrical radiation fields, and so for the IRDCs being studied here (with no internal sources) the alignment would be most efficient on the surfaces of the clouds. Thermal continuum emission from such aligned grains is polarized.
The
field direction can be traced by rotating the polarization vectors by
$90^{\circ}$ \citep{crutcher2012}. Here we present the first such analysis
available for prominent high-mass IRDCs. This data constrain for the
first time the magnetic field properties during the assembly of
massive dense clouds.

\begin{figure*}
  \centerline{\includegraphics[width=\linewidth]{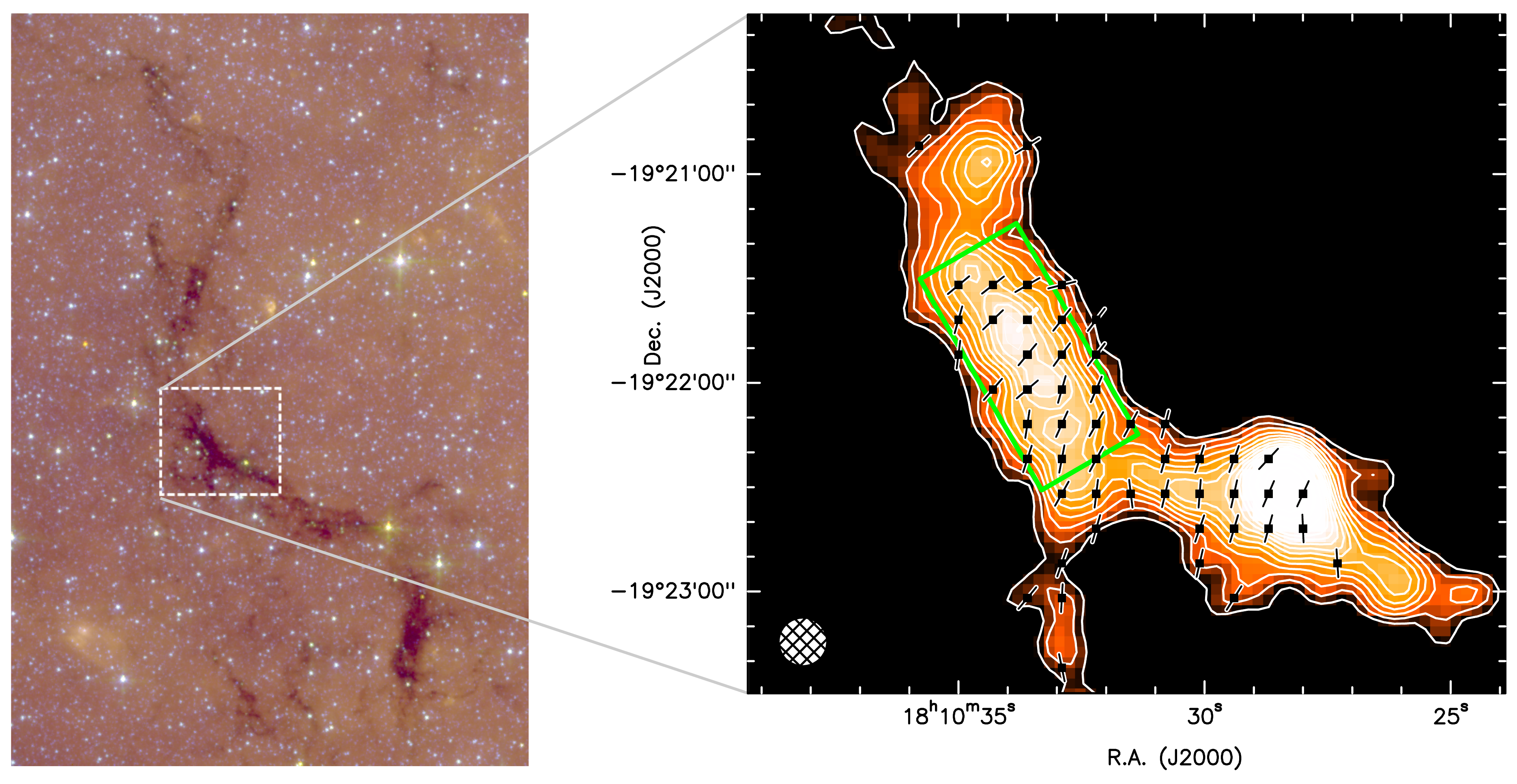}}
  \caption{Polarization data for the G11.11$-$0.12 IRDC.  The \emph{left panel} presents an infrared overview image
    obtained using Spitzer Space Telescope data (at $3.6,5.8,8.0~\rm{}\mu{}m$).  The box indicates the area
    highlighted in the \emph{right panel}. The
    \emph{right panel} shows magnetic field vectors obtained by rotating
    polarization vectors by $90^{\circ}$.  The green box outlines the region for which the
    magnetic field strength is determined. The background and contours
    show SCUBA 850$\mu$m dust intensities.  Contours are drawn in steps of 0.05~Jy/beam,
    starting at 0.1~Jy/beam. Polarization data \citep{Matthews2009} is only shown where
    (\textit{i}) the ratio of the polarization level to its
    uncertainty is $\ge{}3$, corresponding to an error in polarization
    angle $\le{}10^{\circ}$, and (\textit{ii}) the SCUBA 850$\mu$m dust
    intensity is greater than 0.1~Jy/beam. The hatched
    circle corresponds to the SCUBA 850$\mu$m beam. \label{fig:g11}}
\end{figure*}

\begin{figure*}[t]
  \centerline{\includegraphics[width=0.7\linewidth]{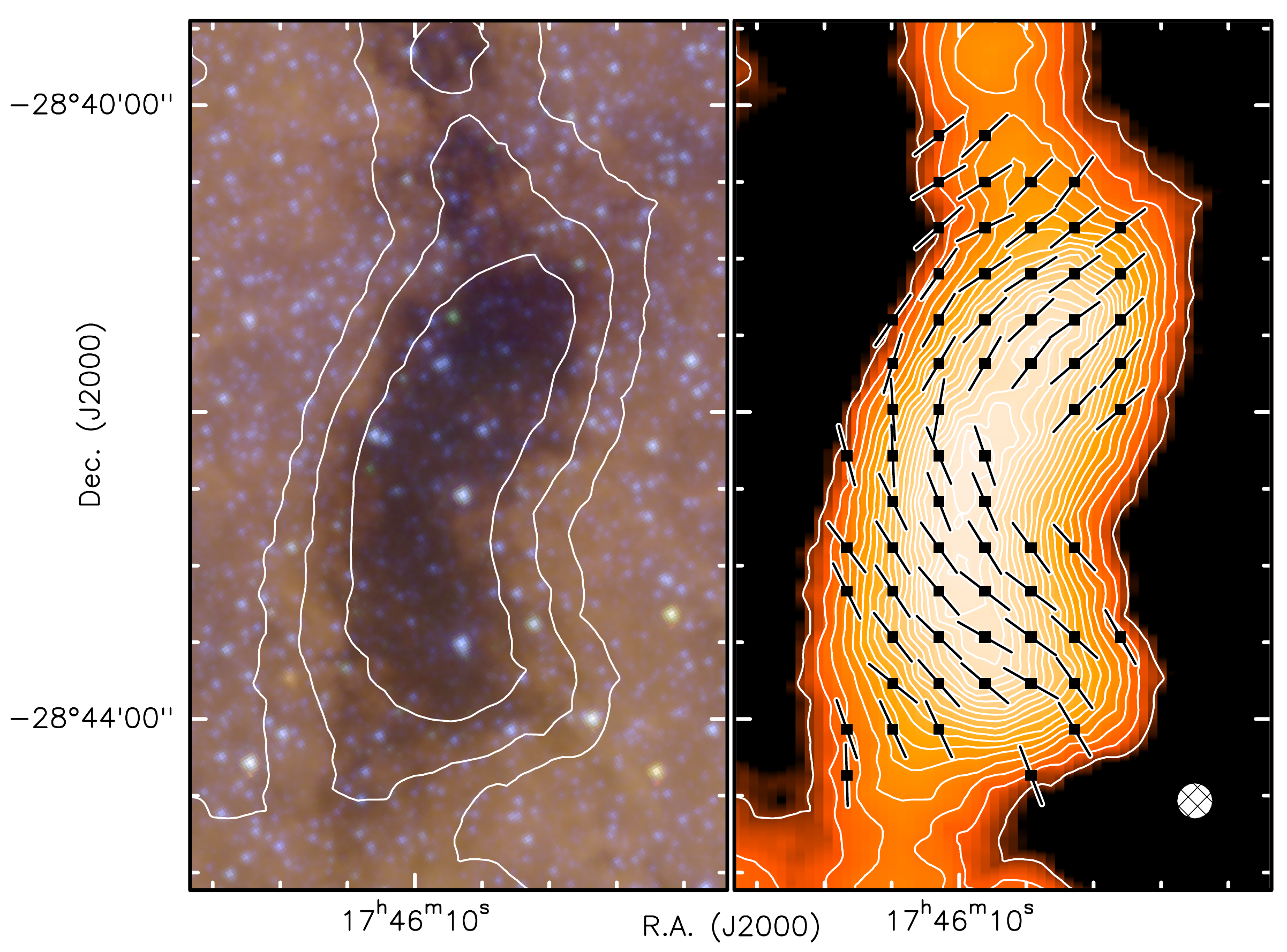}}
  \caption{Polarization data for the G0.253$+$0.016 IRDC. The \emph{left panel} presents an infrared overview image
    obtained using Spitzer Space Telescope data (at
    $3.6,5.8,8.0~\rm{}\mu{}m$).  Selected
    Bolocam 1.1~mm dust intensity \citep{aguirre2011,ginsburg2013}
    contours are overlaid. The
    \emph{right panel} shows magnetic field vectors obtained by rotating
    polarization vectors by $90^{\circ}$.  The background image and
    contours give the Bolocam 1.1~mm dust intensity
    distribution. Contours are drawn in steps of 0.2~Jy/beam,
    starting at 0.2~Jy/beam. Polarization data \citep{Dotson2010} is only shown where
    (\textit{i}) the ratio of the polarization level to its
    uncertainty is $\ge{}3$, corresponding to an error in polarization
    angle $\le{}10^{\circ}$, and (\textit{ii}) the 1.1~mm dust
    intensity is greater than 0.2~Jy/beam. The hatched
    circle corresponds to the CSO 1.1\,mm beam. \label{fig:g0.253}}
\end{figure*}

\section{Observations \& Data Reduction}
\label{sec:observations}

We analyse archival calibrated polarization data obtained  for G11.11$-$0.12. Specifically, we used data for high mass
prestellar cores from the SCUPOL catalog which is a compilation of
calibrated and reduced 850$\mu$m polarization observations made with the JCMT \citep{Matthews2009}. The JCMT project ID for G11.11$-$0.12 is
m03bc32.  Only a part of the 30\,pc long cloud is covered by the
SCUBA polarization observations. 
The data were sampled on a 10\,\arcsec\ pixel grid and the effective
beam width in the map is 20\,\arcsec. Similarly, we use 350$\mu$m archival polarization data from the Caltech
Submillimeter Observatory \citep{Dotson2010} (CSO) for G0.253$+$0.016 which have a  spatial resolution
of 20\,\arcsec. 

\section{Results}
The images presented in Figures~\ref{fig:g11} and ~\ref{fig:g0.253}
summarize the results of the archival polarization observations with
the polarization vectors rotated by $90^{\circ}$ to represent B-vectors, i.e., the orientation of the magnetic field on the plane of the sky at the location of the polarization vector. The
dust continuum emission is shown in colorscale (right panel) with the
plane-of-sky component of B-field (${B}_{\rm{}pos}$) overlaid.  To study the
initial conditions before the onset of star formation, we limit our
analysis of G11.11$-$0.12 to a section of the filament enclosed by the box
in Fig.~\ref{fig:g11}, which is well detached from the $8 \mu$m bright
embedded young high-mass star in the western part of the filament \citep{pillai2006a:g11,gomez2011,ragan2013,wang2014}, while
we study the full extent of G0.253$+$0.016. The mean
position angle (noise-weighted) of the magnetic field for
G11.11$-$0.12, averaged over all the positions shown within the box in
Fig.\,\ref{fig:g11} is 147$^{\circ}$ east of north. On the plane of the sky, the angle
subtended by the main axis of the cloud (i.e. the spine of the filament)
is 30$^{\circ}$ east of north. Therefore, the magnetic field is mainly
perpendicular to the major axis of the filament in
G11.11$-$0.12. The
lowest dust intensity contours also reveal a more diffuse elongated structure
merging onto the main filament, and it is parallel to the magnetic field. For
G0.253$+$0.016, the magnetic field shows a kinked morphology that
correlates well with the structure of the cloud.  The magnetic field is
remarkably ordered in both clouds, with
small angular dispersions $\sigma_{\phi}\le{}17^{\circ}$ relative to the
mean cloud field (see Section~\ref{sec:methods}).

Small angle dispersions suggest very strong fields.  Following
 \citet{chandrasekhar1953}, we may assume that
local perturbations imposed on the mean field direction, characterized
by the standard deviation $\sigma_{\phi}$ of the residual position
angles, are caused by random ``turbulent'' gas motions of
one-dimensional velocity dispersion $\sigma_v$ inside the clouds. Then
 the plane-of-sky component, ${B}_{\rm{}pos}$,  is \citep{chandrasekhar1953}
\begin{equation}
 \label{eq:cf}
   B_{\rm{}pos}=f\,\sqrt{4\pi\varrho}\,\frac{\sigma_v}{\sigma_{\phi}}\,,
\end{equation}
where $\sigma_{\phi}$ is measured in radians and $\varrho$ is the mass density of the region in the cloud relevant to the $\sigma_{\phi}$ and $\sigma_{v}$ values.  We include the correction factor $f=0.5$, based on
  studies using synthetic
  polarization maps generated from numerically simulated
  clouds \citep{ostriker2001,heitsch01}  which suggest that for
  $\sigma_{\phi}\le{}25^{\circ}$ Eq.~(\ref{eq:cf})  is uncertain by
  a factor 2.

\subsection{Methods \label{sec:methods}}

To apply the Chandrasekhar--Fermi method for estimates of the magnetic
field strength, we have to infer how much the magnetic field is disturbed
by the ``turbulent'' gas motions inside the cloud and thus determine
$\sigma_{\phi}$. To do this, we must
essentially separate the magnetic field on the plane of the sky, $\vec{{B}}_{\rm{}pos}$ into an undisturbed
underlying field $\vec{B}_0$ on which disturbances $\vec{B}_{\rm{}t}$
are superimposed such that $\vec{B}=\vec{B}_0+\vec{B}_{\rm{}t}$. For example, in Eq.~\ref{eq:cf}, the
disturbances $\vec{B}_{\rm{}t}$ are characterized via the dispersion of
polarization angles $\sigma_{\phi}$ relative to the direction of the
underlying field. Separating the observed polarization vectors into
the components $\vec{B}_0$ and $\vec{B}_{\rm{}t}$ is the subject of the current subsection.

\subsubsection{Method I: Spatial-Filtering of the Underlying Field}
\medskip

As mentioned earlier, the magnetic field vectors in G11.11$-$0.12
(specifically within the highlighted box in Fig.~\ref{fig:g11}) are roughly aligned
along an axis running South--East to North--West, which is
predominantly perpendicular to the main filament. The dispersion among
these angles shall therefore yield
$\sigma_{\phi}$. In G0.253$+$0.016, the global field structure is, however, more
complex and the underlying field cannot be
approximated as having a spatially constant orientation throughout the
cloud.

A more sophisticated approach is needed to determine a model of
the underlying field that allows for large-scale variations inside the cloud.
A straightforward way of estimating the underlying field is to model the  field as a
distance-weighted mean of the neighboring positions. We thus
subtract the average polarization angle in the neighborhood of every
independent pixel in the map (spatial-filter method).  In G11.11$-$0.12
  no changes in the mean field direction are evident. Thus we use
  filtering scales as large as the size of the box indicated in Fig.~\ref{fig:g11}. In G0.253$+$0.016 the filter should be smaller than the spatial scale on which
  the well--defined background field changes. Inspection of
  Fig.~\ref{fig:g0.253} shows that the field changes its orientation on
  an angular scale $\lesssim{}200\arcsec$. The mean--field removal
  should thus be done on a scale significantly smaller than this. In
  practice we use a maximum filter scale of $200\arcsec/4=50\arcsec$,
  which is significantly larger than the telescope beam. Using data points pre--selected by the spatial
  filter, we then calculate the residual polarization angle as
\begin{equation}
\phi _{i,{\rm{}res}}=\phi _{i}-\frac{\sum_{j=1}^{N}  w_{i,j} \cdot
\phi_{j}}{\sum_{j=1}^{N} w_{i,j}} \,,
\label{eq:spat-filt}
\end{equation}
where  $w_{i,j}=\sqrt{1/S_{i,j}}$ is the weighting function, and
  $S_{i,j}$ is the separation between the pixels. We calculate the dispersion in residual angles
  $\sigma_{\phi,obs}$ as the standard deviation in $\phi_{i,{\rm{}res}}$. 
Even for a perfectly polarized source, noise in the data 
introduces a non-zero dispersion. We then correct the dispersion
for this measurement uncertainty ($\delta_{\phi}$) by subtracting it from the estimated
dispersion, i.e. the corrected dispersion $\sigma_{\phi} =
\sqrt{\sigma_{\phi,obs}^2-\delta_{\phi}^2}$ (see Eqn.\,3 of \citealt{hildebrand2009}).
We find $\sigma_{\phi}$, to be $16\fdg{}2$ for G11.11$-$0.12. This is an upper
  limit to $\sigma_{\phi}$: filter scales smaller than the size of the
 box in Fig.~\ref{fig:g11} yield lower values. In G0.253$+$0.016 a
  maximum filter scale of $200\arcsec/4=50\arcsec$ yields
  $\sigma_{\phi}=9\fdg{}3$. 
Smaller filter scales that are still resolved by the beam yield values as low as
 $8\fdg{}4$, while increasing the
  filter size to $200\arcsec/3=67\arcsec$ gives
  $\sigma_{\phi}=11\fdg{}2$. Thus we adopt $\sigma_{\phi}=9\fdg{}3$
  for G0.253$+$0.016, with an uncertainty of at most 20\%. Results using Method I are reported in Table~1.

\subsubsection{Method II: Structure Function Analysis }
Given the larger number of polarization vectors in G0.253$+$0.016,
we also explore a recent structure function
method \citep{hildebrand2009,houde2009} to test the robustness of our
results. 
This method relies on  the two-point correlation (angular dispersion)
function which has the form
\begin{equation}
\label{eq:houde}
\xi(\ell) \equiv 1-\langle\cos[\Delta{}\Phi(\ell)]\rangle \simeq
\frac{1}{N}
\frac{\langle{}B_{\rm{}t}^2\rangle}{\langle{}B_0^2\rangle}
\left(1-{\rm{}e}^{-\frac{\ell^2}{2(\delta^2+2W^2)}}\right)
+ a_2^\prime\,\ell^2,
\end{equation}
where $\Delta{}\Phi(\ell)$ is the difference in the polarization angle
measured at two positions separated by a distance $\ell$, $B_{\rm{}t}$
and $B_{\rm{}0}$ are the turbulent and large-scale ordered component
respectively, $\delta$ is the turbulent correlation length, $W$ is the
beam radius, and $a_2^\prime$ is the slope of the ``higher-order
effects''. This relation holds when $\ell$ is less than a few times
$W$.  The factor $N$ corrects for the integration along the line of
sight and is defined as the number of independent turbulent cells
probed by observations. Following \citet{houde2009}, if
$\Delta^\prime$ is the depth along the line of sight, then
$N=(\delta^2+2W^2)\Delta^\prime/\sqrt{2 \pi}\delta^3$. The
  magnetic field is then found to have the form \citep{houde2009}
\begin{equation}
  \label{eq:cf2}
   B_{\rm{}0}=\sqrt{4\pi\varrho}\,{\sigma_v}\,\left({\frac{\langle{}B_{\rm{}t}^2\rangle}{\langle{}B_0^2\rangle}}\right)^{-1/2}\,.
\end{equation}
Equations (1) and (4) imply $\langle{}B_{\rm{}t}^2\rangle/\langle{}B_0^2\rangle  \sim [\sigma_{\phi}/ f ]^2$.
The fit
  parameters are $\langle{}B_{\rm{}t}^2\rangle/\langle{}B_0^2\rangle$,
  $\delta$, and $a_2^\prime$, while $W$ and $\Delta^\prime$ are
  assumed.  The fit quality is assessed as the reduced $\chi^2$ by
  comparing the observed two--point correlation function for binned
  separations $\ell_i$, $\xi_{\rm{}obs}(\ell_i)$ to those modelled by
  Eq.~(\ref{eq:houde}), $\xi(\ell_i)$:
\begin{equation}
\chi^2_{\rm{}red} \equiv \frac{\chi^2}{\nu} = \frac{1}{\nu} \,
\sum_i \left( \frac{\xi_{\rm{}obs}(\ell_i) - \xi(\ell_i)}{\sigma(\xi_{\rm{}obs}[\ell_i])} \right)^2 \, ,
\end{equation}
where $\sigma(\xi_{\rm{}obs}[\ell_i])$ is the uncertainty on
$\xi_{\rm{}obs}(\ell_i)$ (Eq.~B6 of \citealt{houde2009}). The number of degrees of
freedom, $\nu$, is given by the number of observations used
in the fit minus the number of free fitting parameters.

We find that the goodness of fit is insensitive to a wide range in values
of $\delta$ when the other two parameters,
$\langle{}B_{\rm{}t}^2\rangle/\langle{}B_0^2\rangle$, and $a_2^\prime$
are left unconstrained. We do therefore follow a different approach to
constrain $\delta$. In the analysis of \citet{houde2009}, $\delta$
describes the spatial scale below which the observed turbulent
fluctuations in velocity and magnetic polarization become small (i.e.,
clouds become ``coherent'' in the terminology of
\citealt{goodman1998:coherent_cores}). Note that this is an
observational property: scales much smaller than the telescope beam
cannot be resolved, and so $\delta\gtrsim{}W$. This conjecture is also
consistent with the hierarchical nature of molecular clouds: observed
properties are always dominated by the structure on the largest
unresolved spatial scale, which is again $\sim{}W$ in our
situation. This implies a total of $N\le{}10$ cells along the line of
sight.  This is consistent with our previous interferometer
observations \citep{kauffmann2013:gc} where we deduce that the cloud
has about 7 velocity components.

In our fits to the data we therefore require $\delta\ge{}W$ as a
constraint to the fit. Only the first four data points are fitted to
fulfill the constraint that separations $\ell\gg{}W$ should be
excluded \citep{houde2009}. The parameter $a_2^\prime$ is left
unconstrained, while we evaluate  $\chi^2$ for a  wide range in the parameter
$\langle{}B_{\rm{}t}^2\rangle/\langle{}B_0^2\rangle$ (see below). This
results in two degrees of freedom ($\nu=2$). Fits consistent with the
data at a confidence level of 90\% must then achieve
$\chi^2_{\rm{}red}<2.3$. Allowing $\delta$ and $a_2^\prime$ to vary as
described above, we find that fixed values of
$\langle{}B_{\rm{}t}^2\rangle/\langle{}B_0^2\rangle\le{}0.5$ are
consistent with this constraint in $\chi^2_{\rm{}red}$. 
This upper limit on $\langle{}B_{\rm{}t}^2\rangle/\langle{}B_0^2\rangle^{1/2} $ is basically an upper limit on $\sigma_{\phi} / f$. If we maintain our estimate of $\sigma_{\phi}$ from Method I, then the result here implies $f \ge 0.23$, instead of $f=0.5$. Based on this we will assume an uncertainty of 2 in Method I associated with Eq.~\ref{eq:cf} where f=0.5 (see Section 3.2).

\subsection{Magnetic Field Strength}
For our calculations we utilize the spatial-filter method (Method I) since it is a model-independent approach that can
be applied to both clouds. 

The density in Eq.~(\ref{eq:cf}) is derived from prior observations of dust
emission. For G11.11$-$0.12, we use the 850$\mu$m dust emission observations reported in
\citet{pillai2006b:nh3} to derive the mass within in
the bounded box shown in Fig.\,~\ref{fig:g11}. For this we  use the temperature 
measurement for the same region \citep{pillai2006b:nh3}. We  approximate the region
within the box as a homogeneous cylinder to estimate an average density. For G0.253$+$0.016, we adopt the average density determined over
the entire cloud (i.e. exactly the same region used in our analysis) from
published mm dust observations \citep{longmore2012:m0.25}. As stated earlier, the alignment mechanism of the dust grains is likely to be most efficient in the outer portions of the cloud and so our determination of $\sigma_{\phi}$ is probably biased towards these outer regions of the cloud. The above derived average densities for G11.11-0.12 and G0.253+0.016 would not be too different from the densities in these outer regions of the clouds, and so these are acceptable estimates of $\rho$ for Eq.~\ref{eq:cf}.

The $\rm{}N_2H^+$ emission shows an excellent correlation with
  dust emission and thus traces the dust throughout the cloud. We
  derive the velocity dispersion from $\rm{}N_2H^+$  observations for
both clouds \citep{pillai:thesis, kauffmann2013:gc}. To do this,
  we spatially integrate all the spectra within the box shown in
  Fig.\,~\ref{fig:g11} for G11.11$-$0.12  and the within the whole cloud shown
  in Fig.\,~\ref{fig:g0.253}. for G0.253+0.016.

It is the non-thermal component of the total velocity dispersion that
influences the MHD turbulence. The  thermal component operates on
scales lower than that of the ambipolar diffusion scale and therefore would
not contribute to the observed dispersion in the field
orientation. Therefore, we equate  $\sigma_{v}$ in Eq.~\ref{eq:cf}
with $\sqrt{ \Delta v_{\rm{} obs}^2 / ({8 \ln 2})-
  k_{\rm{}B}T_{\rm{}g}/m}$, where $\Delta v_{\rm{}obs}$ is the
  line-width observed for the integrated N$_2$H$^+$ line, $k_{\rm{}B}$ and $T_{\rm{}g}$ are
  the Boltzmann constant and the gas temperature, respectively and $m$
  is the mass of a N$_2$H$^+$ molecule. We assume that $\sigma_v$ is
  almost uniform throughout the starless clouds we are studying, so
  the value of $\sigma_v$ derived here is a reasonably good estimate of
  the level of velocity dispersion responsible for the observed
  $\sigma_{\phi}$. Table~1 lists the values
  we use in Eq.~\ref{eq:cf} to determine the plane of the sky
  magnetic field strengths, $B_{\rm{}pos}$ for  G11.11$-$0.12   and
  G0.253+0.016.  Further,
  Table 1 gives the resulting total magnetic field strength for these
  clouds, $B_{\rm{}tot} = 1.3\, B_{\rm{}pos}$, by using an average field geometry
\citep{crutcher04_prestellar}.

Table~1 contains statistical uncertainty estimates obtained via
  regular Gaussian error propagation of noise. In addition, we also consider the
  impact of systematic uncertainties. In our analysis we assume that
  $\sigma_v$ is dominated by it statistical uncertainties only,
    while we assume that
  mass, $N_{\rm{}H_{2}}$ and $\varrho$ is uncertain by a factor 2, while $\sigma_{\phi}$ is
  either an upper limit (for G11.11$-$0.12) or uncertain by a factor
  1.2 (for G0.253+0.016). In addition, we assume that the factor f
    in  Eq.~\ref{eq:cf} has an uncertainty of 2 (see section 3.1.2).
    Eq.~\ref{eq:cf}  essentially means that we must consider a product
    of properties with true values T$_i$ that are scaled by error
    factors r$_i$. If we consider for example a two factor observable,
    T$_{\rm{}obs}$ = T$_1\cdot{} \rm{T}_2$, then a value of
    $T_{\rm{}obs}=r_1T_1\cdot{}r_2T_2$ is within its range of uncertainty.
 Given an uncertainty by a
  factor $a_i$, $r_i$ can vary in the range $1/a_i\le{}r_i\le{}a_i$.
  For our analysis we consider the logarithm of the observed product,
  $\log(T_{\rm{}obs})=\log(T_1)+\log(T_2)+\log(r_1)+\log(r_2)$. We
  assume that the errors $\log(r_i)$ have a flat probability
  distribution within the range limited by $\pm{}\log(a_i)$ and are
  zero for $|\log(r_i)|>\log(a_i)$ (i.e., we adopt a top--hat
  function). The joint probability distribution of
  $\log(r_1)+\log(r_2)$ is then calculated by convolving the
  distributions for the individual $\log(r_i)$. A simple
  transformation back to the linear space allows to use the joint
  probability distribution for $\log(r_1)+\log(r_2)$ to represent the
  probability distribution of $r_1\cdot{}r_2$. We report the resulting
  uncertainties at the 68\% confidence level, corresponding to the
  classical $\pm{}1\cdot{}\sigma$--limit. This approach can without
  any loss of generality be expanded to include an arbitrary number of
  factors $r_iT_i$. 
 In the current example, $B_{pos} \propto T_1T_2T_3T_4 = f
  \varrho^{1/2} \sigma_v \sigma_{\phi}^{-1} $, where a$_1$=2, a$_2 =
  2^{1/2}$, and a$_3 \sim1$ for both clouds. For G0.235+0.016  a$_4$=1.2, and for G11.11-0.12 a$_4$=1, but the upper limit for $\sigma_{\phi}$ for this cloud is tracked through the quantities derived using $\sigma_{\phi}$.

\subsection{Turbulence, Magnetic Field and Gravity}
To assess the importance of turbulence with respect to magnetic
field, we calculate the Alfv\'en Mach number that can be expressed as 
\begin{equation}
\label{eq:MA}
\mathcal{M}_{\rm{}A}=\sqrt{3}\,\sigma_{v}/v_{\rm{}A}
\end{equation}
(see Table~\ref{tab:results}), where
$v_{\rm{}A}=B_{\rm{}tot}/\sqrt{4\,\pi\,\varrho}$ is the Alfv\'en speed
and $B_{\rm{}tot}=1.3\,B_{\rm{}pos}$ converts the projected
plane-of-sky component to the total magnetic field for an average
field geometry \citep{crutcher04_prestellar}. Substitution of
  Eq.~(\ref{eq:cf}) reveals that, except for numerical constants,
  $\mathcal{M}_{\rm{}A}\propto{}\sigma_{\phi}$. The systematic
  uncertainty in the Alfv\'en Mach number is thus identical to the
  small uncertainty in $\sigma_{\phi}$.

 Can self-gravity overcome magnetic forces to initiate collapse in
 high-mass IRDCs? The balance is governed by the mass-to-flux ratio $M/\Phi_B$, where the magnetic
flux $\Phi_B=\pi\,\langle{}B\rangle{}\,R^2$ is derived from the mean
magnetic field and the radius of the cloud cross--section
perpendicular to the magnetic field. The ratio has a critical
value \citep{nakano1978} $(M/\Phi_B)_{\rm{}cr}=1/(2 \pi G^{1/2})$, where
$G$ is the gravitational constant.  Provided
$(M/\Phi_B)<(M/\Phi_B)_{\rm{}cr}$, a cloud will not collapse due to
self--gravity, even if compressed to higher densities, unless
$M/\Phi_B$ increases. We find \citep{mckee:araa07}
\begin{equation}
\frac{(M/\Phi_B)}{(M/\Phi_B)_{\rm{}cr}}=0.76\,\left(\frac{\langle N_{\rm
      H_{2}}\rangle}{10^{23}\,{\rm
      {cm}}^{-2}}\right)\left(\frac{B_{\rm{}tot}}{1000~{}\mu\rm{G}}\right)^{-1}\
\label{eq:mass-to-flux}
\end{equation}
(Table~\ref{tab:results}) when approximating $M$ as
$\pi{}\langle{}N_{\rm{}H_2}\rangle{}R^2$ times the ${\rm{}H_2}$ mass. We assume $\langle{}N_{\rm{}H_2}\rangle$ to be uncertain by
  a factor 2, while the uncertainty of $B_{\rm{}tot}$ comes from
  Sec.~\ref{sec:methods}.

\section{Discussion}
These observations have many important consequences. The first implication is that the magnetic field is dynamically
important relative to turbulence and therefore turbulence is sub-Alfv\'enic. This is
evident from the highly ordered field structure observed in both
G11.11$-$0.12 and G0.253$+$0.016 that is only  slightly perturbed by
turbulence. Analysis of synthetic dust polarization maps generated
from turbulent three-dimensional MHD simulations also show that
sub-Alfv\'enic models show  strongly correlated field lines as
opposed to a very complex structure in super-Alfv\'enic models \citep{falceta-goncalves2008}.
We find Alfv\'en Mach numbers $\mathcal{M}_{\rm{}A} \le 1.2$ 
  given our systematic errors  and formal uncertainties.
This also implies that
simulations of magnetohydrodynamic (MHD) turbulence with
$\mathcal{M}_{\rm{}A}\gg{}1$ should not apply to high-mass stars
formation.

The second consequence is that magnetic fields have a significant
  --- and possibly dominant --- role in shaping the evolution of
  clouds towards gravitational collapse. This follows from upper
  limits ${(M/\Phi_B)}/{(M/\Phi_B)_{\rm{}cr}}\le{}2.1$ including all
  uncertainties, with most likely values
  ${(M/\Phi_B)}/{(M/\Phi_B)_{\rm{}cr}}\lesssim{}1$. These fields may
  dramatically slow down collapse due to self-gravity. The forces due
  to magnetic fields can certainly not be neglected in studies of
  high-mass IRDC evolution.

\citet{myers2009:fils} find that local clouds often form hub-filament
systems. Young clusters form within dense elongated hubs and lower
density filaments converge onto such hubs. Such networks
have been observed in other IRDCs as well
\citep{busquet2013,peretto2013}. Near-infrared polarization
observations that trace magnetic fields in low extinction envelopes of
such filaments have shown that the magnetic field on even larger scale
may be perpendicular to the dense hub
\citep{chapman2011,palmeirim2013, busquet2013,li2014}. The lower
extinction filament in the south of G11.11$-$0.12 that is merging onto
the main dense filament (see Fig.~\ref{fig:g11}) appears to be
consistent with a hub-filament system. While the main dense filament
is perpendicular to the magnetic field, the low column density
filament is parallel to the magnetic field. Therefore, the magnetic field
may funnel gas into the main filament. Higher resolution and
sensitivity mm polarization maps together with information on the
velocity distribution are needed to explore this scenario.
\citet{fiege04:g11} compared the dust continuum data for G11.11$-$0.12
(shown in Fig.~\ref{fig:g11}) to two different magnetic models of
self-gravitating, pressure-truncated filaments which are either
toroidal or poloidal. The observations presented here rule out a
poloidal field. However, they do also not prove the existence of a
toroidal field: the observed field lines might coil around the
filament, but they might as well be relatively straight on spatial
scales larger than the filament and permeate the filament at a random
angle.

In G0.253$+$0.016, the cloud morphology as well as the large
  scale field morphology resemble an arched structure opening to the
  west. Such a morphology might be naively expected when the cloud is
  shocked due to the impact of material approaching from the west. The
  presence of multiple velocity components, as well as the existence
  of spatially extended SiO emission expected in shocks, does indeed
  support such a scenario for G0.253+0.016
  \citep{lis_menten1998:irdc,kauffmann2013:gc}. It is conceivable that
  the impacting material might have been ejected in a supernova. A
  collision between two molecular cloud provides another viable
  explanation. However, there is no unambiguous evidence supporting
  any specific scenario. We do therefore refrain from a detailed
  discussion of the global cloud and magnetic field morphology.

\begin{figure}[t]
\includegraphics[width=0.6\linewidth]{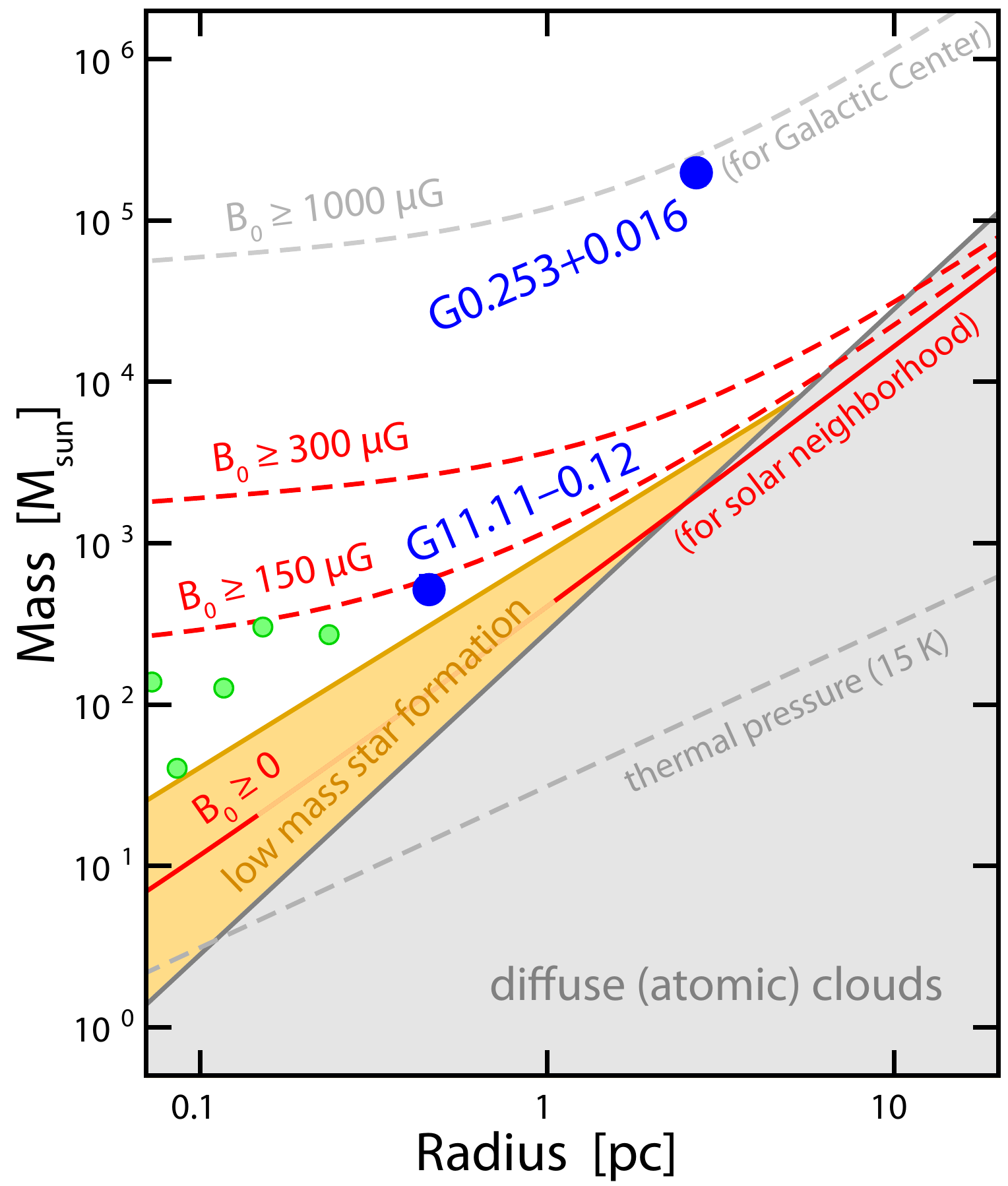}
\caption{Mass versus size relationship for molecular cloud stability. The \emph{solid red line} gives the maximum mass for
 which an unmagnetised cloud of given radius $R$ would be
 stable based on typical
  gas temperatures ($\approx{}15~\rm{}K$) and velocity
 dispersion
 ($\sigma_v\lesssim{}0.8~{\rm{}km\,s^{-1}}\cdot{}[R/{\rm{}pc}]^{0.32}$) in the solar
 neighborhood \citep{kauffmann2013b}. The mass supported by thermal
 pressure is indicated by grey broken line. Diffuse atomic (i.e.,
 non--molecular) clouds, indicated by \emph{grey shading}, and many clouds
 devoid of high-mass star formation, highlighted by \emph{yellow
   shading}, do not need magnetic support to be stabilized against
 collapse. G11.11$-$0.12 and G0.253$+$0.016, indicated by \emph{blue circles}, reside above the red solid
   line, thus requiring significant magnetic support. Potential
   high-mass starless cores \citep{pillai2011a,tan2013}, indicated by \emph{green circles}, are
   also in the mass--size domain requiring significant magnetic
   support.\label{fig:mass-size}}
\end{figure}

The detection of dynamically significant magnetic fields with
$(M/\Phi_B)\lesssim{}(M/\Phi_B)_{\rm{}cr}$ in high-mass dense clouds
possibly resolves a recent riddle in the study of HMSF sites: many of these clouds are so dense and massive that
thermal pressure and random gas motions alone are insufficient to
provide significant support against self--gravity (i.e., these clouds have a
low virial parameter \citep{kauffmann2013b} $\alpha=5\sigma_v^2R/[GM]$). This
suggested \citep{pillai2011a,kauffmann2013b,tan2013} that significant
magnetic fields provide additional support. The data presented here
provide the first direct evidence for this
picture. Figure~\ref{fig:mass-size} illustrates this scenario.  Typical gas temperatures and velocity dispersions are
combined with magnetic fields to provide support against gravitational
collapse. A relation \citep{crutcher2012}
$B=B_0\cdot{}(n_{\rm{}H_2}/10^4~{\rm{}cm^{-3}})^{0.65}$ is adopted, where
$B_0\lesssim{}150~\rm{}\mu{}G$ is common. Many clouds with pure
low--mass star formation can be supported without magnetic fields. But
the more massive HMSF regions need significant magnetic fields, unless
they are in an unlikely \citep{kauffmann2013b} state of rapid collapse. 
 The observed field strengths for G11.11$-$0.12
 and G0.253$+$0.016 (see Table~1) are remarkably consistent with the values
 needed for magnetic support of the two  clouds.

These results indicate that strong magnetic fields may play an important
role in resolving the "fragmentation problem" of high-mass star
formation \citep{krumholz08:1g_cm-2}. Collapsing isothermal molecular clouds would fragment
into a large number of stars with masses of the order of a solar
mass. Increasing the cloud mass would raise the number of stars but
not their mass. High stellar masses, as needed for HMSF, are hard to
realize in this situation. ``Competitive accretion'' models suggest that the stars continue
to grow to higher masses by accreting mass from their
environment \citep{bonnell2001a}. Some ``core accretion \citep{mckee02:100000yrs}'' models propose that the
gas might heat up so that fragmentation is suppressed and
stars of higher mass are formed provided clouds exceed a mass surface
density threshold {\citep{krumholz08:1g_cm-2}
  $\sim{}1~\rm{}g\,cm^{-2}$. Suppression of fragmentation via strong magnetic
fields can provide a natural
explanation without requiring a high surface density
threshold. While fragmentation \citep{butler2012}  and
stability \citep{pillai2011a,kauffmann2013b,tan2013}  studies provide
indirect indications for the presence of strong magnetic fields in
IRDCs, the observations and analyses presented here provide the first
direct evidence. Polarization observations in regions where high-mass
star formation has already initiated reveal that the magnetic field
continues to be  dynamically
significant even after the onset of star formation \citep{girart2013:dr21,zhang2014:bfields}.
 Numerical experiments
with even supercritical fields demonstrate
that the number of fragments is reduced by a factor $\sim{}2$ when
comparing simulations with $(M/\Phi_B)/(M/\Phi_B)_{\rm{}cr}\approx{}2$
to unmagnetised
cases\citep{commercon2011,myers2013:fragmn}. Future
calculations must still show, however, whether this reduction in
fragmentation also results in higher final masses of the stars that
are formed.

The inferred strong magnetic fields can imply a slowing down of the
star formation process. This is the case when
$(M/\Phi_B)/(M/\Phi_B)_{\rm{}cr}\le{}1$: the magnetic flux must be
reduced significantly before collapse and star formation can
occur. Under the influence of strong gravitational forces, ambipolar
diffusion in a medium without random gas motions removes the field in
a period \citep{mckee:araa07}
$\tau_{\rm{}AD}\approx{}1.6\times{}10^6~{\rm{}yr}\,\mathcal{X}_{\rm{}CR}\,(n_{\rm{}H_2}/10^5~{\rm{}cm^{-3}})^{-1/2}$,
where
$\mathcal{X}_{\rm{}CR}=\zeta_{\rm{}CR}/(3\times{}10^{-17}~\rm{}s^{-1})$
captures the impact of the cosmic ray ionization rate
$\zeta_{\rm{}CR}$. This period is long compared to the free--fall
timescale,
$\tau_{\rm{}ff}=9.8\times{}10^4~{\rm{}yr}\,(\langle{}n_{\rm{}H_2}\rangle/10^5~{\rm{}cm^{-3}})^{-1/2}$.
Since $\tau_{\rm{}AD}\sim{}10\,\tau_{\rm{}ff}$, one may thus think
that star formation in magnetically subcritical clouds is very
``slow'' compared to the non--magnetic case. However, clouds must only
loose part of their magnetic flux to become
supercritical \citep{ciolek2000}, and random ``turbulent'' fluctuations
reduce $\tau_{\rm{}AD}$ \citep{fatuzzo2002}.  Even when molecular
  clouds are supported by significant magnetic pressure, ambipolar
  diffusion dictates that star
formation from a self-gravitating core can occur on a timescale only modestly exceeding
$\tau_{\rm{}ff}$.

%%
%% TABLES
%%
%% If there are any tables, put them here.
%%

\begin{sidewaystable}
\caption{Physical Properties and Magnetic Field
  Parameters\label{tab:results}}
\hspace*{-2cm} \begin{minipage}{\linewidth}
\begin{center}
{\footnotesize
\begin{tabular}{cccccccccccccc}
\hline \hline
Source & Distance  &  $\sigma_{\rm{}obs}(v)$  & Mass  & $\langle{}N_{\rm{}H_{2}}\rangle$ & Density  &  $\sigma_{\phi}$ &
$\hbox{B}_{\rm tot}$  &  $\mathcal{M}_{\rm{}A}$ & $\frac{(M/\Phi_B)}{(M/\Phi_B)_{\rm{}cr}}$\\
  & pc & $\mathrm{km\,s^{-1}}$ & $\hbox{M}_\odot$  & $10^{23}\,{\rm{cm}}^{-2}$ &   10$^{4}$\,$\mathrm{cm^{-3}}$ & deg. & $\mu$G & \\ \hline
G11.11$-$0.12$^1$   & 3600  & $0.9\pm{}0.1$ & $549\rvert_{275}^{1098}$
& $0.4 \rvert_{0.2}^{0.8}$ & $3\rvert_{2}^{6}$ & $<16.2\pm{}1.6$ & $>267\pm{}26 \rvert_{163}^{437}$ & $<0.8\pm{}0.1
\rvert_{0.5}^{1.2}$ & $<1.1\pm{}0.1 \rvert_{0.6}^{2.1}$ \\
G0.253$+$0.016     & 8400  & $6.4\pm{}0.4$  &
$(2\rvert_{1}^{4})\times10^5 $ & $4 \rvert_{2}^{8}$  & $8
\rvert_{4}^{16}$  & $9.3\pm{}0.9 \rvert_{7.8}^{11.2}$ & $5432\pm{}525
\rvert_{3312}^{8908}$ &  $0.4\pm{}0.1\rvert_{0.3}^{0.7}$  &
$0.6\pm{}0.1\rvert_{0.3}^{1.1}$
\end{tabular}
}
\end{center}

\end{minipage}\\
Notes:   gas velocity dispersion ($\sigma_{\rm{}obs}(v)$) \citep{pillai:thesis,kauffmann2013:gc}, 
 average column density over the
same area where $\hbox{B}_{\rm pos}$ is measured ($\langle{}N_{\rm{}H_{2}}\rangle$),
H$_2$ density (Density),\citep{pillai2006b:nh3,longmore2012:m0.25}, standard deviation of the residual
polarization angles ($\sigma_{\phi}$),
 total magnetic field ($\hbox{B}_{\rm tot}$), Alfv\'en mach number
($\mathcal{M}_{\rm{}A}$), mass-to-flux parameter
$({M/\Phi_B})/{(M/\Phi_B)_{\rm{}cr}}$. Note that we use a molecular
weight per hydrogen molecule of 2.8 to convert from mass to particle
density \citep{kauffmann2008}.\\
$^1$Estimates are made for the material within
the bounded box shown in Figure~\ref{fig:g11}. \\
There are two distinct sources of errors in our calculations:
statistical and systematic.  A $\pm$ sign is adopted to show the statistical
uncertainty determined
from a Gaussian error propagation. Since our systematic uncertainties are asymmetric
about the central value, we provide the minimum and the maximum
value with in a  68\% central confidence region (i.e $1 \sigma$) about
the central estimate. This is represented by the lower and upper bound
respectively for the relevant parameters.
\end{sidewaystable}

\acknowledgements{We thank P. Redman for his data on G11.11$-$0.12 and his discussion on the observations. We thank Brenda Matthews
for kindly checking the quality of the re-processed SCUPOL data for
this source. We thank the referee for a very constructive review of
the manuscript which improved the quality of the manuscript. }

Facilities: \facility{JCMT (SCUBA, SCUPOL)}, \facility{CSO
  (Hertz, Bolocam)}, \facility{Spitzer (GLIMPSE)}


\begin{thebibliography}{51}
\expandafter\ifx\csname natexlab\endcsname\relax\def\natexlab#1{#1}\fi

\bibitem[{{Aguirre} {et~al.}(2011){Aguirre}, {Ginsburg}, {Dunham}, {Drosback},
  {Bally}, {Battersby}, {Bradley}, {Cyganowski}, {Dowell}, {Evans}, {Glenn},
  {Harvey}, {Rosolowsky}, {Stringfellow}, {Walawender}, \&
  {Williams}}]{aguirre2011}
{Aguirre}, J.~E., {Ginsburg}, A.~G., {Dunham}, M.~K., {et~al.} 2011, \apjs,
  192, 4

\bibitem[{{Bonnell} {et~al.}(2001){Bonnell}, {Bate}, {Clarke}, \&
  {Pringle}}]{bonnell2001a}
{Bonnell}, I.~A., {Bate}, M.~R., {Clarke}, C.~J., \& {Pringle}, J.~E. 2001,
  \mnras, 323, 785

\bibitem[{{Busquet} {et~al.}(2013){Busquet}, {Zhang}, {Palau}, {Liu},
  {S{\'a}nchez-Monge}, {Estalella}, {Ho}, {de Gregorio-Monsalvo}, {Pillai},
  {Wyrowski}, {Girart}, {Santos}, \& {Franco}}]{busquet2013}
{Busquet}, G., {Zhang}, Q., {Palau}, A., {et~al.} 2013, \apjl, 764, L26

\bibitem[{{Butler} \& {Tan}(2012)}]{butler2012}
{Butler}, M.~J. \& {Tan}, J.~C. 2012, \apj, 754, 5

\bibitem[{{Carey} {et~al.}(1998){Carey}, {Clark}, {Egan}, {Price}, {Shipman},
  \& {Kuchar}}]{carey1998:irdc}
{Carey}, S.~J., {Clark}, F.~O., {Egan}, M.~P., {et~al.} 1998, \apj, 508, 721

\bibitem[{{Chandrasekhar} \& {Fermi}(1953)}]{chandrasekhar1953}
{Chandrasekhar}, S. \& {Fermi}, E. 1953, \apj, 118, 116

\bibitem[{{Chapman} {et~al.}(2011){Chapman}, {Goldsmith}, {Pineda}, {Clemens},
  {Li}, \& {Kr{\v c}o}}]{chapman2011}
{Chapman}, N.~L., {Goldsmith}, P.~F., {Pineda}, J.~L., {et~al.} 2011, \apj,
  741, 21

\bibitem[{{Ciolek} \& {Basu}(2000)}]{ciolek2000}
{Ciolek}, G.~E. \& {Basu}, S. 2000, \apj, 529, 925

\bibitem[{{Commer{\c c}on} {et~al.}(2011){Commer{\c c}on}, {Hennebelle}, \&
  {Henning}}]{commercon2011}
{Commer{\c c}on}, B., {Hennebelle}, P., \& {Henning}, T. 2011, \apjl, 742, L9

\bibitem[{{Crutcher}(2012)}]{crutcher2012}
{Crutcher}, R.~M. 2012, \araa, 50, 29

\bibitem[{{Crutcher} {et~al.}(2004){Crutcher}, {Nutter}, {Ward-Thompson}, \&
  {Kirk}}]{crutcher04_prestellar}
{Crutcher}, R.~M., {Nutter}, D.~J., {Ward-Thompson}, D., \& {Kirk}, J.~M. 2004,
  \apj, 600, 279

\bibitem[{{Dotson} {et~al.}(2010){Dotson}, {Vaillancourt}, {Kirby}, {Dowell},
  {Hildebrand}, \& {Davidson}}]{Dotson2010}
{Dotson}, J.~L., {Vaillancourt}, J.~E., {Kirby}, L., {et~al.} 2010, \apjs, 186,
  406

\bibitem[{{Falceta-Gon{\c c}alves} {et~al.}(2008){Falceta-Gon{\c c}alves},
  {Lazarian}, \& {Kowal}}]{falceta-goncalves2008}
{Falceta-Gon{\c c}alves}, D., {Lazarian}, A., \& {Kowal}, G. 2008, \apj, 679,
  537

\bibitem[{{Fatuzzo} \& {Adams}(2002)}]{fatuzzo2002}
{Fatuzzo}, M. \& {Adams}, F.~C. 2002, \apj, 570, 210

\bibitem[{{Fiege} {et~al.}(2004){Fiege}, {Johnstone}, {Redman}, \&
  {Feldman}}]{fiege04:g11}
{Fiege}, J.~D., {Johnstone}, D., {Redman}, R.~O., \& {Feldman}, P.~A. 2004,
  \apj, 616, 925

\bibitem[{{Ginsburg} {et~al.}(2013){Ginsburg}, {Glenn}, {Rosolowsky},
  {Ellsworth-Bowers}, {Battersby}, {Dunham}, {Merello}, {Shirley}, {Bally},
  {Evans}, {Stringfellow}, \& {Aguirre}}]{ginsburg2013}
{Ginsburg}, A., {Glenn}, J., {Rosolowsky}, E., {et~al.} 2013, \apjs, 208, 14

\bibitem[{{Girart} {et~al.}(2013){Girart}, {Frau}, {Zhang}, {Koch}, {Qiu},
  {Tang}, {Lai}, \& {Ho}}]{girart2013:dr21}
{Girart}, J.~M., {Frau}, P., {Zhang}, Q., {et~al.} 2013, \apj, 772, 69

\bibitem[{{G{\'o}mez} {et~al.}(2011){G{\'o}mez}, {Wyrowski}, {Pillai},
  {Leurini}, \& {Menten}}]{gomez2011}
{G{\'o}mez}, L., {Wyrowski}, F., {Pillai}, T., {Leurini}, S., \& {Menten},
  K.~M. 2011, \aap, 529, A161

\bibitem[{{Goodman} {et~al.}(1998){Goodman}, {Barranco}, {Wilner}, \&
  {Heyer}}]{goodman1998:coherent_cores}
{Goodman}, A.~A., {Barranco}, J.~A., {Wilner}, D.~J., \& {Heyer}, M.~H. 1998,
  \apj, 504, 223

\bibitem[{{Heitsch} {et~al.}(2001){Heitsch}, {Zweibel}, {Mac Low}, {Li}, \&
  {Norman}}]{heitsch01}
{Heitsch}, F., {Zweibel}, E.~G., {Mac Low}, M.-M., {Li}, P., \& {Norman}, M.~L.
  2001, \apj, 561, 800

\bibitem[{{Henning} {et~al.}(2010){Henning}, {Linz}, {Krause}, {Ragan},
  {Beuther}, {Launhardt}, {Nielbock}, \& {Vasyunina}}]{henning2010:g11}
{Henning}, T., {Linz}, H., {Krause}, O., {et~al.} 2010, \aap, 518, L95

\bibitem[{{Hildebrand} {et~al.}(2009){Hildebrand}, {Kirby}, {Dotson}, {Houde},
  \& {Vaillancourt}}]{hildebrand2009}
{Hildebrand}, R.~H., {Kirby}, L., {Dotson}, J.~L., {Houde}, M., \&
  {Vaillancourt}, J.~E. 2009, \apj, 696, 567

\bibitem[{{Houde} {et~al.}(2009){Houde}, {Vaillancourt}, {Hildebrand},
  {Chitsazzadeh}, \& {Kirby}}]{houde2009}
{Houde}, M., {Vaillancourt}, J.~E., {Hildebrand}, R.~H., {Chitsazzadeh}, S., \&
  {Kirby}, L. 2009, \apj, 706, 1504

\bibitem[{{Kainulainen} {et~al.}(2013){Kainulainen}, {Ragan}, {Henning}, \&
  {Stutz}}]{kainulainen2013}
{Kainulainen}, J., {Ragan}, S.~E., {Henning}, T., \& {Stutz}, A. 2013, \aap,
  557, A120

\bibitem[{{Kauffmann} {et~al.}(2008){Kauffmann}, {Bertoldi}, {Bourke}, {Evans},
  \& {Lee}}]{kauffmann2008}
{Kauffmann}, J., {Bertoldi}, F., {Bourke}, T.~L., {Evans}, II, N.~J., \& {Lee},
  C.~W. 2008, \aap, 487, 993

\bibitem[{{Kauffmann} {et~al.}(2013{\natexlab{a}}){Kauffmann}, {Pillai}, \&
  {Goldsmith}}]{kauffmann2013b}
{Kauffmann}, J., {Pillai}, T., \& {Goldsmith}, P.~F. 2013{\natexlab{a}}, \apj,
  779, 185

\bibitem[{{Kauffmann} {et~al.}(2013{\natexlab{b}}){Kauffmann}, {Pillai}, \&
  {Zhang}}]{kauffmann2013:gc}
{Kauffmann}, J., {Pillai}, T., \& {Zhang}, Q. 2013{\natexlab{b}}, \apjl, 765,
  L35

\bibitem[{{Krumholz} \& {McKee}(2008)}]{krumholz08:1g_cm-2}
{Krumholz}, M.~R. \& {McKee}, C.~F. 2008, \nat, 451, 1082

\bibitem[{{Lazarian}(2007)}]{lazarian2007}
{Lazarian}, A. 2007, \jqsrt, 106, 225

\bibitem[{{Li} {et~al.}(2014){Li}, {Goodman}, {Sridharan}, {Houde}, {Li},
  {Novak}, \& {Tang}}]{li2014}
{Li}, H.-b., {Goodman}, A., {Sridharan}, T.~K., {et~al.} 2014, ArXiv e-prints

\bibitem[{{Lis} \& {Menten}(1998)}]{lis_menten1998:irdc}
{Lis}, D.~C. \& {Menten}, K.~M. 1998, \apj, 507, 794

\bibitem[{{Longmore} {et~al.}(2012){Longmore}, {Rathborne}, {Bastian}, {Alves},
  {Ascenso}, {Bally}, {Testi}, {Longmore}, {Battersby}, {Bressert}, {Purcell},
  {Walsh}, {Jackson}, {Foster}, {Molinari}, {Meingast}, {Amorim}, {Lima},
  {Marques}, {Moitinho}, {Pinhao}, {Rebordao}, \&
  {Santos}}]{longmore2012:m0.25}
{Longmore}, S.~N., {Rathborne}, J., {Bastian}, N., {et~al.} 2012, \apj, 746,
  117

\bibitem[{{Matthews} {et~al.}(2009){Matthews}, {McPhee}, {Fissel}, \&
  {Curran}}]{Matthews2009}
{Matthews}, B.~C., {McPhee}, C.~A., {Fissel}, L.~M., \& {Curran}, R.~L. 2009,
  \apjs, 182, 143

\bibitem[{{McKee} \& {Ostriker}(2007)}]{mckee:araa07}
{McKee}, C.~F. \& {Ostriker}, E.~C. 2007, \araa, 45, 565

\bibitem[{{McKee} \& {Tan}(2002)}]{mckee02:100000yrs}
{McKee}, C.~F. \& {Tan}, J.~C. 2002, \nat, 416, 59

\bibitem[{{Myers} {et~al.}(2013){Myers}, {McKee}, {Cunningham}, {Klein}, \&
  {Krumholz}}]{myers2013:fragmn}
{Myers}, A.~T., {McKee}, C.~F., {Cunningham}, A.~J., {Klein}, R.~I., \&
  {Krumholz}, M.~R. 2013, \apj, 766, 97

\bibitem[{{Myers}(2009)}]{myers2009:fils}
{Myers}, P.~C. 2009, \apj, 700, 1609

\bibitem[{{Nakano} \& {Nakamura}(1978)}]{nakano1978}
{Nakano}, T. \& {Nakamura}, T. 1978, \pasj, 30, 671

\bibitem[{{Ostriker} {et~al.}(2001){Ostriker}, {Stone}, \&
  {Gammie}}]{ostriker2001}
{Ostriker}, E.~C., {Stone}, J.~M., \& {Gammie}, C.~F. 2001, \apj, 546, 980

\bibitem[{{Palmeirim} {et~al.}(2013){Palmeirim}, {Andr{\'e}}, {Kirk},
  {Ward-Thompson}, {Arzoumanian}, {K{\"o}nyves}, {Didelon}, {Schneider},
  {Benedettini}, {Bontemps}, {Di Francesco}, {Elia}, {Griffin}, {Hennemann},
  {Hill}, {Martin}, {Men'shchikov}, {Molinari}, {Motte}, {Nguyen Luong},
  {Nutter}, {Peretto}, {Pezzuto}, {Roy}, {Rygl}, {Spinoglio}, \&
  {White}}]{palmeirim2013}
{Palmeirim}, P., {Andr{\'e}}, P., {Kirk}, J., {et~al.} 2013, \aap, 550, A38

\bibitem[{{Perault} {et~al.}(1996){Perault}, {Omont}, {Simon}, {Seguin},
  {Ojha}, {Blommaert}, {Felli}, {Gilmore}, {Guglielmo}, {Habing}, {Price},
  {Robin}, {de Batz}, {Cesarsky}, {Elbaz}, {Epchtein}, {Fouque}, {Guest},
  {Levine}, {Pollock}, {Prusti}, {Siebenmorgen}, {Testi}, \&
  {Tiphene}}]{perault1996:iso}
{Perault}, M., {Omont}, A., {Simon}, G., {et~al.} 1996, \aap, 315, L165

\bibitem[{{Peretto} {et~al.}(2013){Peretto}, {Fuller}, {Duarte-Cabral},
  {Avison}, {Hennebelle}, {Pineda}, {Andr{\'e}}, {Bontemps}, {Motte},
  {Schneider}, \& {Molinari}}]{peretto2013}
{Peretto}, N., {Fuller}, G.~A., {Duarte-Cabral}, A., {et~al.} 2013, \aap, 555,
  A112

\bibitem[{{Pillai}(2006)}]{pillai:thesis}
{Pillai}, T. 2006, PhD thesis, Max-Planck-Institut f{\"u}r Radioastronomie

\bibitem[{{Pillai} {et~al.}(2011){Pillai}, {Kauffmann}, {Wyrowski}, {Hatchell},
  {Gibb}, \& {Thompson}}]{pillai2011a}
{Pillai}, T., {Kauffmann}, J., {Wyrowski}, F., {et~al.} 2011, \aap, 530, A118+

\bibitem[{{Pillai} {et~al.}(2006{\natexlab{a}}){Pillai}, {Wyrowski}, {Carey},
  \& {Menten}}]{pillai2006b:nh3}
{Pillai}, T., {Wyrowski}, F., {Carey}, S.~J., \& {Menten}, K.~M.
  2006{\natexlab{a}}, \aap, 450, 569

\bibitem[{{Pillai} {et~al.}(2006{\natexlab{b}}){Pillai}, {Wyrowski}, {Menten},
  \& {Kr{\"u}gel}}]{pillai2006a:g11}
{Pillai}, T., {Wyrowski}, F., {Menten}, K.~M., \& {Kr{\"u}gel}, E.
  2006{\natexlab{b}}, \aap, 447, 929


\bibitem[{{Ragan} {et~al.}(2013){Ragan}, {Henning}, \& {Beuther}}]{ragan2013}
{Ragan}, S.~E., {Henning}, T., \& {Beuther}, H. 2013, ArXiv e-prints

\bibitem[{{Tan} {et~al.}(2014){Tan}, {Beltran}, {Caselli}, {Fontani}, {Fuente},
  {Krumholz}, {McKee}, \& {Stolte}}]{tan2014}
{Tan}, J.~C., {Beltran}, M.~T., {Caselli}, P., {et~al.} 2014, ArXiv e-prints

\bibitem[{{Tan} {et~al.}(2013){Tan}, {Kong}, {Butler}, {Caselli}, \&
  {Fontani}}]{tan2013}
{Tan}, J.~C., {Kong}, S., {Butler}, M.~J., {Caselli}, P., \& {Fontani}, F.
  2013, \apj, 779, 96

\bibitem[{{Wang} {et~al.}(2014){Wang}, {Zhang}, {Testi}, {Tak}, {Wu}, {Zhang},
  {Pillai}, {Wyrowski}, {Carey}, {Ragan}, \& {Henning}}]{wang2014}
{Wang}, K., {Zhang}, Q., {Testi}, L., {et~al.} 2014, \mnras

\bibitem[{{Zhang} {et~al.}(2014){Zhang}, {Qiu}, {Girart}, {Hauyu}, {Liu},
  {Tang}, {Koch}, {Li}, {Keto}, {Ho}, {Rao}, {Lai}, {Ching}, {Frau}, {Chen},
  {Li}, {Padovani}, {Bontemps}, {Csengeri}, \& {Juarez}}]{zhang2014:bfields}
{Zhang}, Q., {Qiu}, K., {Girart}, J.~M., {et~al.} 2014, ArXiv e-prints

\end{thebibliography}
\end{document}